\documentclass{svjour3}                     % onecolumn (standard format)
\smartqed  % flush right qed marks, e.g. at end of proof
\usepackage{graphicx}
\begin{document}

\title{Geometric model of the structure of the neutron%\thanks{Grants or other notes
}


\author{Jacek Syska
}


\institute{J. Syska \at
              Department of Field Theory and Particle Physics, Institute of Physics,\\
University of Silesia, Uniwersytecka 4, 40-007 Katowice, Poland,  and \\
              Departamento de F{\'\i}sica Te\'orica y del Cosmos and
              CAFPE, Universidad de Granada, E-18071 Granada, Spain
\\
              \email{jacek.syska@us.edu.pl}           %  \\
}

\date{Received: date / Accepted: date}

\maketitle

\begin{abstract}{The paper examines the geometrical properties of
a six-dimensional Kaluza-Klein type model. They may have an impact
on the model of the structure of a neutron and its excited states
in the realm of one particle physics. The statistical reason for
the six-dimensionality and the stability of the solution is given.
The derivation of the weak limit approximation of the general wave
mechanical (quantum mechanical) approach, defined in the context
of losing its self-consistency (here gravitational), is presented.
The non self-consistent case for the Klein-Gordon equation is
defined. The derivation of the energy of states and the analysis
of the spin origin of the analyzed fields configuration is
presented as the manifestation of both the geometry of the
internal two-dimensional space and kinematics of fields inside it.
The problem of the departure from the (gravitational)
self-consistent calculations of the metric tensor and of other
fields of the configuration is discussed. The implementation of
the model for the description of a neutron and its excited states,
including their spins and energies, is given. The informational
reason for the existence of the internal extra space dimensions is
proposed.} 
\end{abstract}

%
%\PACS{       {14.20.Dh}{wave mechanics of neutron}   \and
%  {04.90.+e }{self-consistent Kaluza-Klein field theory }
%     } % end of PACS codes} %end of abstract
%
%\maketitle
%
\section{Introduction}
\label{introduction}

The paper presents the model of an elementary particle and
particularly the model of the neutral nucleon and its excited
states. Since the model is atomic-like I will give a concise
reminder both of the history of the models of an atom and of a
neutron which still overlap until the present day.

Possibly the first model of an atom in the 20-th century was
Kelvin's static one \cite{Kelvin} of a sphere of uniformly
distributed positive electricity with embedded discrete electrons,
followed closely by the Thomson's model of "the pudding with
raisin muffins in it" \cite{Thomson1904}. Models also appeared at
the end of the 19-th century, stimulated by empirical facts
suggesting the complexity of the structure of the atom and its
composition from smaller components \cite{Thomson_Chemical
Rubber_Co}. As a result, both the theoretical intuition and the
empirical indications prompted the sense of the electron to
Thomson. He disagreed with the idea of the planetary "point
planets" model of the atom but probably was influenced by
Maxwell's considerations on Saturn's rings \cite{Maxwell}. Hence,
contrary to Kelvin's model, the one by Thomson \cite{Thomson1904}
is not static and one of its characteristics is the motion of a
ring of {\it n} negatively charged particles (electrons
\cite{Thomson1897}, as they were called by Stoney \cite{Stoney}),
which are set within an uniformly charged sphere (with the radius
of the order of $10^{-10}$ m) having a number of particles which
vary from ring to ring. The attractive force of the positively
charged sphere and the repulsion of the negatively charged
electrons arranged in a series of parallel rings were to guarantee
the stability of the system.  The strange characteristic of the
model is that if the number of the negative particles is larger
than 5, some of them must be set in the central position of the
sphere yet not necessarily strictly in the center, or else the
movement in the atom becomes unstable. This problem still arouses
interest until the present day \cite{Luca-Rodrigues-Levin}.  \\
Thomson's atom has two types of vibrations: the first one
connected with the movement of electrons along their orbits and
the second one developing out of the distortion of a ring from its
equilibrium circular shape in a particular spherical shell.
Although the model does not explain the characteristic
regularities of the observed spectral lines and the deflection of
an electron when it passes through matter, it was regarded as a
sufficient one by many physicists until the end of the first
decade of the 20th century. This was due to the fact that he found
a close relation between the occupation of the electrons inside
rings formed in the concentric shells and the regularities inside
the Mendeleev's periodic table. However, at the same time some
scientists considered that the planetary model of the atom was
more appropriate. \\
In 1920 Rutherford considered his model of the neutron in relation
to the compressed hydrogen atom in the core of a star and
presented his hypothesis on the structure of the neutron
\cite{Rutherford}. He perceived the neutron as an electron pressed
into the interior of a proton hence his model of the neutron was
more in line with Thomson's idea of the atom than of the planetary
kind. In 1932 Chadwick experimentally confirmed Rutherford's
hypothesis on the existence of the neutron.\\
In the period from 1925 to 1928 there emerged three formulations
of quantum mechanics: Heisenberg's \cite{Heisenberg}, Dirac's
\cite{Dirac} and Schr\"{o}dinger's \cite{Schrodinger}. They were
developed nearly at the same time. The Schr\"{o}dinger equation
provides a direct representation of the mechanical system using
the notion of the wave function for a particular state. Here the
eigenstate functions form the orthogonal reference frame in the
Hilbert space in which the wave function and its time evolution
can be analyzed. The specific Schr\"{o}dinger's substantial
approach to the square modulus of the wave function gave the name
'wave mechanics' to his interpretation of quantum mechanics which
is in line with de Broglie's theory of phase waves
\cite{Broglie,Espinosa 1982}.  \\
Two questions about the main characteristics of the type of the
models of an atom or nucleon built in these formalisms arise.
Firstly, are they Thomson's with deep overlapping of constituent
fields or planetary-like, for which the constituent fields do not
overlap so deeply? On the surface, all of them carry some
planetary features, especially if the Hamiltonians of the
particular models are analyzed. This may be easily understood
because in this respect they are classical mechanics descendants,
yet in their main idea connected with the nonlocality of the wave
function they are Thomson oriented, regardless of the fact whether
the interpretation of the wave function is Schr\"{o}dinger's or
Copenhagen's. Indeed, in the atom the electron wave function is
deeply embedded in the nucleus field.  Secondly, what is the
geometry of the manifold (space or space-time) which forms the
background arena for the disputed configurations of fields? The
common feature of Thomson's and Schr\"{o}dinger's like models of
the atom is the homogeneity of their space-time structure. We will
come back to this problem in the passage below.\\
Now, let us return to the neutron. Since it seemed that quantum
mechanics did not permit a consistent representation of the
neutron as a bound state of the proton and electron, the before
mentioned Rutherford's conception of the structure of the neutron
has been rejected by
%Pauli, Fermi, Heisenberg, and
the majority of physicists. Yet, until now attempts have been made
to restore this model. For instance such attempts are connected
with the construction of a covering of quantum mechanics known as
hadronic mechanics \cite{Santilli} for the specific objective of
achieving a consistent quantitative treatment of nonlocal,
nonlinear and nonpotential effects in deep wave-overlappings of
particles at short distances. As far as the quantum models of the
neutron are considered, the most popular is the quantum field
theory version of the bag model. Setting aside the difficulties of
this model with the determination of the spin of the nucleon we
give some literature references \cite{bag} only citing after
\cite{Heyde}: "... all spin parts $\left[ {\rm of \; the \;
nucleon} \right] $ have to add to $\frac{1}{2}$ which is
incredible in the light of the present day experiments. This may
indicate that some underlying symmetries, unknown at present, are
playing a role in forming the various contributing parts such that
the final sum rule gives the fermion $\frac{1}{2}$ value".\\
Another story is partly connected with the name of Riemann. In his
lecture entitled "On the hypotheses that lie at the foundations of
geometry" \cite{Riemann}, given
%on 10-th June of
in 1854, he emphasized that the truth about space has to be
discovered\footnote{I do not agree with this point of view, but
definitely the truth about space is consistent with physical
experience. I consider the physical world to be perhaps at least
six-dimensional with at least four of these dimensions to be of
the space-time origin.} from physical experience. The clue is that
he noted that the geometry of space could be highly irregular at
very small distances, yet appearing smooth at the observed ones.
He wrote: "Space [in the large] if one ascribes to it a constant
curvature, is necessarily finite, provided only that this
curvature has a positive value, however small...~. It is quite
conceivable that the geometry of space in the very small does not
satisfy the
axioms of [Euclidean] geometry...~".\\
But the story of the atom and nucleon may be told and retold (at
least up to now) in many ways \cite{Lakhtakia}. Not that all of
them are right. The latest attractive model of the (hydrogen)
atom, based on the isomorphism between Maxwell and Dirac
formalisms (or rather on optics-mechanics isomorphism called
Hamilton's analogy), is connected with the name of Sallhofer
\cite{Sallhofer}. He worked out the formal mathematical strong
similarity (I would not call it identity), in the Minkowski space,
between electrodynamics and wave mechanics by means of which he
proved that the hydrogen atom might be seen as a pair of mutually
refracting electromagnetic waves. Previously this similarity was
pointed out with amusement by Sakurai \cite{Sakurai_2}.
%as a footnote.

Below the model of a neutral nucleon will be presented which is
more planetary (fields do not overlap very deeply) and
incorporates the Riemann's idea of space. The metric of the
underlying space-time was previously obtained in
\cite{Dar,Dziekuje Ci Panie Jezu Chryste} where a six -
dimensional Kaluza -- Klein type model at the classical level is
considered. In them the static spherically symmetric solution to
the coupled six-dimensional Einstein and Klein-Gordon equations
was derived in the presence of the basic massless dilaton field
$\varphi$ which forms a kind of ground field for the self-field of
gravitation, the notions of which were discussed elsewhere
\cite{bib B-K-1,Dziekuje_Jacek_nova_1,Dziekuje_Jacek_nova_2}. The
solutions presented in \cite{Dar} are parameterized by parameter
$A$ which has similar dynamical consequences as mass $M = A c^2/(2
G)$ \cite{Dziekuje za galaktyka}, i.e. its existence would be
perceived by an observer in the same way as invisible mass which
could be the extended "center" of a particle. Because the solution
is horizon free hence it is fundamentally different from the four
- dimensional Schwarzschild one.

\subsection{The motivations for extra dimensions}
\label{Fisher}

The idea of the six-dimensional space-time re-enters the physics
occasionally \cite{Nishino-Sezgin}-\cite{Manka-Syska-a}. Yet the
motivations for choosing the six-dimensional models were diverse.
Foe instance in \cite{Nishino-Sezgin,Salam-Sezgin-1984} Nishino,
Salam and Sezgin suggested that one may obtain the fermion
spectrum in four-dimensions within the framework of D=6, $N=2$
Kaluza-Klein supergravity. A six-dimensional model of the
Kaluza-Klein theory was also previously investigated by Ma{\'n}ka
and Syska \cite{Manka-Syska-a,Dziekuje_Jacek_nova_1}, and by
Ivashchuk, Melnikov and Bronnikov
\cite{Iv-Mel-1994,Iv-Mel-1995,Br-Mel-1995}. Recently Sparling
\cite{Sparling} has followed the concept of $(3+3)$ dimensional
structure of the space-time \cite{Demers}. Unfortunately, models
described in \cite{Sparling,Demers} contain the time component
with two additional time dimensions.

However there is a way to extend the four-dimensional space-time
to the multi-dimensional one via the Fisherian statistical
analysis. Few years ago Frieden proved \cite{Frieden} that the
Fisher information channel capacity is the statistical ancestor of
the kinematical part of the well known field theory models. Using
the channel capacity notion, which is inherently connected with
the internal parametric space of the sample (collected by the
particle alone), and two informational principles, the structural
and the variational one, Frieden obtained the Klein-Gordon
equation of motion with the proper relativistic dispersion
relation\footnote{He obtained the proper structural equation of
motions for the Maxwell electromagnetic field, for the Dirac field
and for the gravitational field also \cite{Frieden}. }
\cite{Frieden}. The structural information principle has been
recently proven in \cite{Dziekuje_informacja}. The point is that
the channel capacity which has turned out to be the kinematical
part of the action of any field theory model is additive both in
the Minkowskian $\nu=0,1,2,3$ and the sample $n=1,2,...,\aleph$
(parametric) indices \cite{Frieden}. This enables the partition of
the channel capacity $I$ into two parts, one for the
four-dimensional space-time and the other for the inner parametric
space of the sample whose dimension depends on the spinorial
representation of the field \cite{Frieden,Dziekuje_informacja}.
When we note by $x^{\nu}_{n} = y^{\nu}_{n} - \theta^{\nu}_{n}$ the
added fluctuations of the data $y^{\nu}_{n}$ (collected by the
system alone) from the expectation positions $\theta^{\nu}_{n}
\equiv \theta^{\nu}x_{n}$ \cite{Frieden}, then it can be proven
that it takes the following form:
\begin{eqnarray}  \label{Fisher_information}
I &=& 4  \sum_{n=1}^{\aleph} \int \! d^{4}y_{n} \! \left(
\frac{\partial q_{n}(y^{\nu}_{n})}{\partial \theta^{\nu}_{n}}
\frac{\partial q_{n}(y^{\nu}_{n})}{\partial \theta_{\nu n}}
\right) \\
&=& \int \! d^{4}x \, {\cal L}_{4} \left(g_{\mu \nu}, \,
\varphi_{4}(x^{\nu}) \right) + \int \! d^{\aleph}x \, {\cal
L}_{\aleph} \left( g_{m n}, \, \varphi_{\aleph}(x_{n}) \right) \ ,
\nonumber
\end{eqnarray}
where $q_{n}(y^{\nu}_{n})$ are the original field amplitudes in
the sample\footnote{After squaring each one of
$q_{n}(y^{\nu}_{n})$ and multiplication, the likelihood of the
sample is calculated.}
\cite{Frieden,Dziekuje_informacja,Amari_Nagaoka} and $x^{\nu}_{n}$
are the original Fisherian variables. The amplitudes are
factorized as follows $q_{n}(y^{\nu}_{n}) = \varphi_{4}(x^{\nu})
\, \varphi_{\aleph}(x_{n})$ and the metric tensor $g_{MN}$ ($M,N
\equiv (\mu,\nu=0,...,3; m,n=1,...,\aleph)$) is obtained
effectively from this procedure \cite{Denisov-Logunov}. Factor $4$
in the kinematical form of $I$ signifies the Fisher information
origin of the action, yet it does not enter into the equation of
motion \cite{Frieden} as it is also factored out from the
structural information \cite{Dziekuje_informacja}. \\ For the
complex scalar field the dimension $\aleph$ of the extra
parametric space is equal to $\aleph=2$. In this way the
six-dimen\-sionality of the space having the mixed space-time and
parametric character, is chosen. Hence the model with the internal
space geometry of the two-dimensional torus parameterized by two
angles will be presented. This provides us with geometry of the
internal space, which in Sections~3 and 4 enables the description
of the spinorial field. In \cite{Dar} Biesiada, Ma{\'n}ka and
Syska showed that the six-dimensionality of this space-time
enables the self-consistency of the background solution of the
coupled Einstein and Klein-Gordon equations with internal space,
compactified in a non-homogeneous manner. For the sake of clarity,
the summary of \cite{Dar} is presented  below in Section~2. In
Sections~3 and 4 the model will raise a new issue on the
wave-mechanical analysis of the possible general structure of a
particle. In Section~4 its application to the description of the
neutron and its excited states is presented. In general,
statistical models which one can obtain in this way are of the
Kaluza-Klein type with four-dimensional space-time and
$\aleph$-dimensional internal (parametric) space.

\section{The geometry of the space-time}
\label{geometry}

In \cite{Dar,Dziekuje Ci Panie Jezu Chryste} a $(1+5)$
six-dimensional field theory has been considered (see also
\cite{Chen}) which comprises the gravitational self field
described by a metric tensor, $g_{MN}$, and a real massless
"basic" scalar (dilatonic) field, $\varphi$. We have decomposed
the action into two parts:
\begin{eqnarray}  \label{row_dzialanie-EH-fi}
{\cal S}  = {\cal S}_{EH} + {\cal S}_{\varphi} = \!\! \int \! d^{
6}x  \sqrt{- g} \, \frac{1}{2 \kappa_{6}} \, {\cal R} + \int \!
d^{ 6}x \sqrt{- g} \ \frac{- g_{MN}}{2} \,
\partial^{M} \varphi \, \partial^{N} \varphi \,  ,
\end{eqnarray}
where $S_{EH}$ is the Einstein --- Hilbert action and
$S_{\varphi}$ is the action for a real massless scalar (dilatonic)
field with the Lagrangian density equal to ${\cal L}_{\varphi} = -
\frac{1}{2} \: g_{MN} \: \partial^{M} \varphi \, \partial^{N}
\varphi$ .
%This scalar field $\varphi$ is the created fluctuation out from nothingness.
%\begin{eqnarray}  \label{row_2}
%{\cal L}_{\varphi} = - \frac{1}{2} \: g_{MN} \:
%\partial^{M} \varphi \partial^{N} \varphi \; .
%\end{eqnarray}
In Eq.(\ref{row_dzialanie-EH-fi}) $g = det\; g_{MN}$ denotes the
determinant of the metric tensor, ${\cal R}$ is the curvature
scalar of the six - dimensional (in general curved) space-time,
and $\kappa_{6}$ denotes the coupling constant of the
six-dimensional theory, analogous to the familiar Newtonian
gravity constant.\\
By extremalizing the action given by
Eq.(\ref{row_dzialanie-EH-fi}) we obtain the Einstein equations
\begin{equation}  \label{row_3}
G_{MN} = \kappa_{6} \: T_{MN},
\end{equation}
where $G_{MN} = R_{MN} - \frac{1}{2} \ g_{MN} {\cal R}\,$ is the
Einstein tensor, $R_{MN}$ is the six - dimensional Ricci tensor,
${\cal R}$ is the curvature scalar and $T_{MN}$ is the energy -
momentum tensor of a real scalar (dilatonic) field $\varphi$ which
is given by
\begin{equation}  \label{row_4}
T^{M}_{\;\, N} = \partial_{N} \varphi \: \frac{\partial {\cal
L}_{\varphi}} {\partial (\partial_{M} \varphi )} - \delta^{M}_{\;
N} {\cal L}_{\varphi} \; .
\end{equation}
Variation of the total action ${\cal S}$ with respect to the field
$\varphi$ gives the Klein - Gordon equation
\begin{equation}  \label{row_5}
\qed \varphi \equiv - \frac{1}{\sqrt{- g}} \: \partial_{M}
(\sqrt{-g}\; g^{MN} \partial_{N}) \varphi = 0 \;\;  ,
\end{equation}
where $g^{MN}$ is the tensor dual to $g_{MN}$.

Now consider the six - dimensional space-time which is a
topological product of the curved four - dimensional physical
space-time (with the metric $g_{\alpha \omega }, \; \alpha ,
\omega = 0,1,2,3 $) and the internal compactified space (with the
metric $g_{he}, \; h,e = 5,6 $). Therefore the metric tensor can
be factorized as
\begin{eqnarray}  \label{row_6}
g_{MN} = \pmatrix{g_{\alpha \omega } & 0 \cr 0 & g_{he} } \; .
\end{eqnarray}
The four-dimensional diagonal part is assumed to be that of a
spherically symmetric geometry
\begin{eqnarray}  \label{row_7}
g_{\alpha \omega} = \pmatrix{e^{\nu (r)} & & & \cr & - e^{\mu (r)}
& 0 & \cr & 0 & - r^{2} & \cr & & & - r^{2} sin^{2}\Theta \cr } \;
,
\end{eqnarray}
where $\nu (r)$ and $\mu (r)$ are (at this stage) two arbitrary
functions. Analogously, we take the two-dimensional internal part
to be
\begin{eqnarray}  \label{row_8}
g_{he} = \pmatrix{ - \varrho^{2} (r) \: cos^{2} \vartheta & 0 \cr
0 & - \varrho^{2} (r) } \; .
\end{eqnarray}
The six-dimensional coordinates $(x^{M})$ are denoted by $(t, r,$
$\Theta, \Phi, \vartheta, \varsigma)$ where $t \in [0,\infty )$ is
the usual time coordinate, $r\in [0,\infty ),$ $\Theta \in [0,\pi
] $ and $\Phi \in [0,2 \pi )$ are the familiar three-dimensional
spherical coordinates in the macroscopic space; $\vartheta \in
[-\pi,\pi)$
%\bigcup_{k \in Z}[- \pi /2 + 2 k \pi , 3/2 \pi + 2 k \pi]
and $\varsigma \in [0, 2 \pi )$ are coordinates in the internal
two-dimensional parametric space and $\varrho \in (0,\infty )$ is
the ``radius'' of this internal space. We assume that $\varrho (r)
$ is the function of the radius $r$ in the external
three-dimensional space. The internal space is a 2-dimensional
topological torus with $r$-dependent parameter $\varrho (r)$.
Using Eqs.(\ref{row_7})-(\ref{row_8}), we can calculate the
components of the Ricci tensor. They are given in Appendix~A.

Now we seek for a solution of the Einstein equations (see
Eq.(\ref{row_3})) with the Ricci tensor given in Appendix~A by
Eqs.(\ref{row_10})-(\ref{row_14}), with $\nu (r) = \mu(r)$, and
with the following boundary conditions
\begin{eqnarray}  \label{row_16}
\lim_{r\rightarrow \infty } \nu (r) = \lim_{r\rightarrow \infty }
\mu (r) = 0 \; , \;\;\; \lim_{r\rightarrow \infty } \varrho (r) =
d = constant \neq 0 \; ,
\end{eqnarray}
which at the spatial infinity reproduces the flat external
four-dimensional Minkowski space-time and the static internal
space of ``radius'' $d$ which is of the order calculated in
Section~4. We also suppose that the scalar field $\varphi$ is the
function of the radius $r$ alone, $\varphi = \varphi (r)$, and we
impose the following boundary condition for the scalar field
$\varphi$:
\begin{equation}  \label{row_18}
\lim_{r\rightarrow \infty } \varphi (r) = 0 \;
\end{equation}
which supplements boundary conditions (\ref{row_16}) for the
metric components.

By virtue of Eqs.(\ref{row_4}) and (\ref{row_dzialanie-EH-fi}) it
is easy to see that the only nonvanishing components of the
energy-momentum tensor are
\begin{eqnarray}  \label{row_19}
\!\!\!\!\! - \; T^{r}_{r} = T^{t}_{t} = T^{\Theta}_{\Theta} =
T^{\Phi}_{\Phi} = T^{\vartheta}_{\vartheta} =
T^{\varsigma}_{\varsigma} = \frac{1}{2} \: g^{rr} \: (\partial_{r}
\varphi )^{2} .
\end{eqnarray}

Consequently, it is easy to verify that the self-consistent
solution of the coupled Einstein (\ref{row_3}) and Klein-Gordon
(\ref{row_5}) equations is
\begin{eqnarray}  \label{row_20}
\nu (r) = \mu (r) = \ln \left( \frac{r}{r + A} \right) \; ,
\;\;\;\; \varrho (r) =d \: \sqrt{\frac{r + A}{r}}
\end{eqnarray}
\begin{eqnarray}  \label{row_22}
\varphi (r) = \pm \sqrt{\frac{1}{2 \kappa_{6}}} \:\ln \left(
\frac{r}{r + A} \right) \; .
\end{eqnarray}
Hence it results that the only nonzero component of the Ricci
tensor (see Eqs.(\ref{row_10})-(\ref{row_14}) in Appendix~A) is
$R^{r}_{r}$. So the curvature scalar ${\cal R}$ is equal to
\begin{eqnarray}  \label{row_24}
{\cal R} = R^{r}_{r} = \frac{A^{2}}{2 \: r^{3} (r + A)} \; ,
\end{eqnarray}
where $A$ is the real constant, with the dimensionality of length,
whose value is to be taken from the observation of each particular
system (but see Section~4). We notice that all of the six diagonal
Einstein equations (\ref{row_3}) have shrunk to just one
\begin{eqnarray}  \label{row_25}
\frac{1}{2} \: {\cal R} = \kappa_{6} \: T^{r}_{r} \; .
\end{eqnarray}
Now we can rewrite the metric tensor in the form
\begin{eqnarray}  \label{row_26}
\!\!\!\!\! g_{MN} = diag \, (\frac{r}{r + A},\, - \frac{r}{r +
A},\,
 - r^{2},\, - r^{2} sin^{2}\Theta, \, -  d^{2} \frac{r + A}{r} cos^{2} \vartheta,\, - d^{2} \frac{r
+ A}{r} )
\end{eqnarray}
with its determinant equal to
\begin{eqnarray}  \label{row_27}
g = det g_{MN} = - (d^{2} \: r^{2} \: sin\Theta \: cos\vartheta
)^{2} \; .
\end{eqnarray}
To summarize we notice that the real massless "basic" free scalar
field $\varphi (r)$ (see Eq.(\ref{row_22})) can be the source of
the nonzero metric tensor as in Eq.(\ref{row_26}). Only when the
constant $A$ is equal to zero, the solutions (\ref{row_20}) --
(\ref{row_22}) become trivial and the six-dimensional space-time
is Ricci flat.

It is worth noting that since (with the solutions given by
Eqs.(\ref{row_20}) and (\ref{row_22})) the components
$R^{\vartheta}_{\vartheta}$ and $R^{\varsigma}_{\varsigma}$ of the
Ricci tensor (Appendix~A) are equal to zero for all values of $A$,
the internal space is always Ricci flat. However, we must not
neglect the internal space because its ``radius'' $\varrho$ is a
function of $r$ and the two spaces, external and internal, are
therefore ``coupled''. Only when $A = 0$ are these two spaces
``decoupled'', and the four-dimensional space-time becomes
Minkowski flat\footnote{When $A$ is not equal to zero, our
four-dimensional external space-time is curved. Its scalar
curvature ${\cal R}_{4}$ can be given by Eq.(\ref{row_24}), i.e. $
{\cal R}_{4} = {\cal R} = \frac{A^{2}}{2 \: r^{3} (r + A)}$ . }.

\subsection{Stability of the background
solution} \label{stability}

The self consistent solution given by Eqs.(\ref{row_22}) and
(\ref{row_26}) of the coupled Einstein and Klein-Gordon equations
is unique. To answer the question on the stability of this
self-consistent gravity-dilatonic configuration let us calculate
its energy:
\begin{eqnarray}
\label{config energy} E_{g + \varphi} = \int_{V} d^{5} x \ \sqrt{-
g} \ \left( G^{tt} + \kappa_{6} \ T^{tt} \right)
\end{eqnarray}
where using Eqs.(\ref{row_19})-(\ref{row_27}) we obtain:
\begin{eqnarray}
\label{conig energy value} E_{g + \varphi} = - 2 \int_{V} d^{5} x
\ \sqrt{- g} \, g^{tt} \, \frac{\cal R}{2} = - 2 \, Q \;
\end{eqnarray}
with
\begin{eqnarray}
\label{Q} Q = 8\,d^2\,{\pi }^2 \, \lim_{\varepsilon \rightarrow
0^{+}}\int _{\varepsilon}^{\infty} \frac{A^2}{r^2}\,dr  = (2 {\pi
}\, d) \, ( 4\,{\pi }\, A^{2}) \lim_{\varepsilon \rightarrow
0^{+}} \frac{d}{\varepsilon}  \, .
\end{eqnarray}
The integral of $1/r^2$ does not converge on $(0, \infty)$, hence
the self consistent solution (\ref{row_22}) and (\ref{row_26})
cannot be obtained in the limit $E_{g + \varphi} \rightarrow 0$,
from the convergent solutions around $E_{g + \varphi} = 0$
corresponding to $A = 0$. \\ If we depart from the self consistent
solution then a similar result can be obtained from the
information analysis in the following way. From the notion of the
Fisher information \cite{Frieden}, we notice that $d$ is (for $r
\rightarrow \infty$, see Eq.(\ref{row_20})) the characteristic
dimension of the inner parametric space, i.e. the radius of the
expected value of the position of the system. In the Frieden
analysis $r$ is the added fluctuation to $d$ \cite{Frieden}. Hence
if we assume that the fluctuation $r$ is not smaller than the
expected value $d$ then the physical limit for the fluctuation $r$
is $\varepsilon \approx d$.  In Eq.(\ref{Q}) the cutoff
$\varepsilon = d$ can be taken which leads to $Q = Q_{cut} = (2
{\pi }\, d) \, (4\,{\pi }\, A^{2})$. Therefore $Q_{cut}$ is finite
and its value could be even small in contradistinction to the
infinite value obtained in Eq.(\ref{Q}) for $\varepsilon
\rightarrow 0$ in the self consistent case.  Hence, we see that
the unique self consistent solution given by Eqs.(\ref{row_22})
and (\ref{row_26}) cannot be destabilized to yield any other.
  \\
Let us also notice that in Eq.(\ref{config energy}) the partition
of $E_{g + \varphi}$ into two parts can be obtained by using the
Fisher information formalism developed for physical models by
Frieden \cite{Frieden}. According to \cite{Frieden} $G^{tt}$ is
connected with the Fisherian kinematical degrees of freedom of the
gravitational configuration
whereas $T^{tt}$ with its structural degrees of freedom. \\
To conclude this Section it is worth noting that the metric given
by Eq.(\ref{row_26}), with the dilatonic field given by
Eq.(\ref{row_22}), serves, under further conditions discussed
below, as a background fields configuration. This metric appears
in the equation of motions for all new fields which  weakly enter
the system.

\section{The quantum implication - Klein-Gordon equation.
Wave-mechanical approach and weak interaction limit}
\label{implication}

Until now the calculations were fully self-consistent. The model
presented below adds a new scalar field to the system. Yet, since
it procures big analytical complications the model will stop being
fully self-consistent. This means among others that we will
decline from making the self-consistent correction of the metric
(\ref{row_26}) received in Section~2. Therefore the following
calculations are made totally on the basis of the wave-mechanical
approach.

Let us investigate the Klein--Gordon relativistic wave equation
for the motion of a scalar particle with a wave function $\phi$
and mass $m$ (for $m^{2} < 0$ it would be a $6$-dimensional
tachion),
\begin{equation}
\label{rownanie} \frac{1}{\sqrt{- g}} \: (i \hbar)
\partial_{M} \left( \sqrt{-g}\; g^{MN} (i \hbar) \, \partial_{N} \,
\phi \right) - m^{2} c^{2} \phi = 0 \, ,
\end{equation}
in the six - dimensional space-time given by the central
gravitational field described by the metric, $g_{MN}$, given by
Eq.(\ref{row_26}) with its determinant given by Eq.(\ref{row_27}).
This metric tensor is used in the above Klein--Gordon equation as
the background metric only. Hence we neglect the modification of
the Einstein equations. Therefore a particle described by the
$\phi$ wave function moves without changing the metric $g_{MN}$ in
reciprocal action. This is called the wave-mechanical approach.
{\it But to make our calculations self-consistent by the inclusion
of the Einstein equations coupled to Klein-Gordon equation, we
should follow the pattern of the previous section for all scalar
fields (particles) which have the mass $m$ and are inherently tied
up by strong interaction (compare Section~\ref{bounding_energy}).
This procedure should give the self-consistent change of the
metric tensor $g_{MN}$.}

We incorporate the so-called "natural interpretation" of the wave
equation\footnote{\label{f-cyt-12} Then we do not have any
problems with possible negative values of probability density
which is the inherent property of the probabilistic interpretation
of quantum mechanics for the Klein--Gordon equation
\cite{Schiff-book}.}
\cite{Dziekuje_Jacek_nova_1,Dziekuje_Jacek_nova_2} with a particle
as an oscillating substance described by the wave function $\phi$
which is deformable according to Eq.(\ref{rownanie}).\\
Using Eqs.(\ref{row_26})~and~(\ref{row_27}), we can rewrite the
Klein-Gordon equation as follows:
\begin{eqnarray}
\label{K-G-m} & & \!\!\!\!\!\!\! sin\Theta \, cos\vartheta \,
\left( \frac{r + A}{r} r^{2} \, \frac{\partial}{c \partial t} (
\frac{\partial \phi} {c
\partial t}) -
\frac{\partial}{\partial r}
 (r^{2} \, \frac{r + A}{r} \, \frac{\partial \phi}{\partial r})
 \right)
\nonumber \\
& & \!\!\!\!\!\!\! - \ r^{2} cos\vartheta \left(
\frac{\partial}{\partial \Theta} (sin\Theta \frac{1}{r^{2}} \,
\frac{\partial \phi} {\partial \Theta})  +  sin\Theta \,
\frac{1}{r^{2} \,
sin^{2} \Theta} \, (\frac{\partial^{2} \phi}{\partial \Phi^{2}}) \right) \nonumber \\
& & \!\!\!\!\!\!\! - \ \frac{r^{2}}{d^{2}} sin\Theta \left(
\frac{r}{r + A} \, \frac{\partial}{\partial \vartheta}
(cos\vartheta \, \frac{\partial \phi}{\partial \vartheta})  +
cos\vartheta
\frac{r}{r + A} \, \frac{\partial^{2} \phi}{\partial \varsigma^{2}} \right) \nonumber \\
&  & \!\!\!\!\!\!\! + \ r^{2}  sin\Theta \, cos\vartheta \,
\frac{m^{2} \, c^{2}} {\hbar^{2}} \, \phi = 0 \; .
\end{eqnarray}
In order to isolate the time dependence, the standard procedure of
the separation of variables is performed with the following
factorization of $\phi$
\begin{equation}
\label{3} \phi ({\bf r}, \vartheta, \varsigma, t) = u({\bf r},
\vartheta, \varsigma) e^{-i \frac{E t}{\hbar}} \; .
\end{equation}
In this way, we obtain the set of stationary states. As the result
of further separation in the three-dimensional spherical
coordinates ${\bf r} =(r,\Theta,\Phi) $ of the macroscopic space
and in $(\vartheta ,\varsigma)$ coordinates of the internal
two-dimensional space described by the factorization
\begin{equation}
\label{4} u({\bf r}, \vartheta, \varsigma) = u_{r} (r) \, Y_{LM}
(\Theta, \Phi) \, y_{l \tilde{\mathrm{m}}} (\vartheta, \varsigma)
\; ,
\end{equation}
we obtain the "almost" familiar radial Klein-Gordon wave equation
\cite{Schiff-book}
\begin{eqnarray}
\label{5} & & \!\!\!\!\!\!\!\!\!\!\!\! \left( \frac{r}{r + A}
\right) \biggl\{ - \frac{1}{r^{2}} \frac{ \partial}{\partial r}
\left( (\frac{r + A}{r}) r^{2}\, \frac{\partial u_{r}}{\partial r}
\right) + \frac{L
(L + 1)}{r^{2}} \, u_{r} \biggr\} \nonumber \\
& & \!\!\!\!\!\!\!\!\!\!\!\!\!\! = \frac{1}{\hbar^{2} c^{2}}
\biggl\{ E^{2} - m^{2} c^{4} (\frac{r}{r + A}) - \lambda_{d}
\frac{\hbar^{2} c^{2}}{d^{2}} \left( \frac{r}{r + A} \right)^{2}
\biggr\}  \, u_{r}
\end{eqnarray}
with the angular equation in the macroscopic space which defines
states with a definite three--dimensional angular momentum
\begin{eqnarray}
\label{6} &-& \hbar^{2} \left[ \frac{1}{sin\Theta}
\frac{\partial}{\partial \Theta} \left( sin\Theta
\frac{\partial}{\partial \Theta} \right) + \frac{1}{sin^{2}
\Theta} \, \frac{\partial^{2}}{\partial \Phi^{2}} \right] Y_{LM}
\nonumber \\ &=& L (L + 1) \hbar^{2} Y_{LM} \; ,
\end{eqnarray}
and the angular equation in the inner space with a definite
internal angular momentum
\begin{eqnarray}
\label{7} \!\!\!\!\!\!\! \frac{1}{cos\vartheta}
\frac{\partial}{\partial \vartheta} \left( cos\vartheta
\frac{\partial}{\partial \vartheta} y_{l \tilde{\mathrm{m}}}
\right) + \frac{1}{\partial \varsigma} \left( \frac{\partial
}{\partial \varsigma} y_{l \tilde{\mathrm{m}}} \right) = -
\lambda_{d} \, y_{l \tilde{\mathrm{m}}} \, ,
\end{eqnarray}
where $\lambda_{d}$ is the main internal angular quantum number.
In Eq.(\ref{6}) functions $Y_{LM}$ are the spherical harmonics
(normalized angular momentum eigenfunctions) with $L = 0,1,2,...$
and $M = 0, \pm 1,..., \pm L$. Similarly $y_{l
\tilde{\mathrm{m}}}$ are the periodic functions of the inner
angles $\vartheta$ and $\varsigma$. As the result of further
separation of Eq.(\ref{7}) in the internal coordinates described
by the factorization
\begin{equation}
\label{8} y_{l \tilde{\mathrm{m}}} (\vartheta, \varsigma) = g_{l}
(\vartheta) h_{\tilde{\mathrm{m}}} (\varsigma) \; ,
\end{equation}
we obtain
\begin{eqnarray}
\label{9} \frac{1}{cos \vartheta} \frac{\partial}{\partial
\vartheta} \left( cos \vartheta \frac{\partial}{\partial
\vartheta} g_{l} \right) + k_{l} \, g_{l} = 0  \, ,
\end{eqnarray}
where $k_{l}$ is the internal $\vartheta$-angle quantum number of
the torus geometry (see Section~\ref{further}).

\subsection{The case with $\lambda_{d} = 0$ and $m = 0$ and its difference from others}
\label{global}

It is not difficult to verify that for $\lambda_{d} = 0$ and $m =
0$ the only solution of the stationary Klein-Gordon equation
Eq.(\ref{5}) alone, vanishing in infinity (although not
normalizable), exists for $E = 0$ and $L = 0$. For the radial part
of $\phi$ this solution has the following form (see
Eq.(\ref{row_22}) for $\varphi (r)$):
\begin{eqnarray}
\label{Pcd7} u_{r}(r) =  C \, \ln \left( \frac{r}{r + A} \right)
\; ,
\end{eqnarray}
where $C$ is a constant. In this way we formally obtain \cite{Dar}
up to $C$ the same solution which follows from the classical
Klein-Gordon equation for the dilaton field (see
Eq.(\ref{row_5})). Yet now the field is a scalar one, hence the
energy of the configuration (\ref{Pcd7}) is positive and equals:
\begin{eqnarray}
\label{T00} \int_{V} d^{5}x \, \sqrt{- g} \, T^{tt}_{\phi} = 2 \,
C^{2} \, Q \;\, ,
\end{eqnarray}
where $T^{tt}_{\phi}$ is the $tt$ component of the energy-momentum
tensor of the scalar field $\phi$. Integral $Q$ does not converge
and is given by Eq.(\ref{Q})\footnote{As before in Eq.(\ref{Q}),
the integral of $1/r^2$ in $Q$ does not converge on $(0, \infty)$.
Hence the solution (\ref{Pcd7}) obtained for $E = 0$, $m = 0$,
$\lambda_{d} = 0$ and $L = 0$ cannot be obtained from the
solutions around $E = 0$, $m = 0$ (i.e. in the limit $E
\rightarrow 0$, $m \rightarrow 0$) for which the L{\small HS} of
Eq.(\ref{T00}) converges.}.
%where $C = \frac{1}{\sqrt{2 \kappa_{6}}}$.
Yet, fortunately here, we may recall the self-consistent
calculations for fields which were previously introduced, i.e. the
dilaton $\varphi$ and metric $g_{MN}$ ones, with the now added
scalar field $\phi$, which we for the time being assume to be
given by the configuration (\ref{Pcd7}). The coupled Einstein and
Klein-Gordon field equations for all these fields have to be
solved. Now, the case with $\lambda_{d} = 0$, $m = 0$, $E =0$ is
the only one for which the value of $\hbar$ in Eq.(\ref{5}) is
irrelevant as then the R{\small HS} of Eq.(\ref{5}) becomes equal
to zero. Therefore in this case the energy-momentum tensor for the
massless scalar field enters into the R{\small HS} of Einstein
equations (\ref{row_3}) in the same way as the dilaton field,
differing only in sign. Hence the R{\small HS} of Einstein
equations (\ref{row_3}) becomes equal to zero leading to Ricci
flat solution of Einstein equations. Therefore the discussed new
configuration of all fields, $g_{MN}$, $\varphi$ and $\phi$, does
not describe the gravitational bound state, contrary to the case
of the metric-dilaton field configuration given by
Eqs.(\ref{row_22}) and (\ref{row_26}) alone. Hence the
configuration given by Eq.(\ref{Pcd7}) for $m=0$ and $\lambda_{d}
= 0$ does not appear\footnote{We may say that solution
(\ref{Pcd7}) of Eq.(\ref{5}) is in this case as "macroscopic" as
the dilatonic configuration (\ref{row_22}) leading to the zeroing
on the R{\small HS} of the Einstein equations (\ref{row_3}). In
other words: The only Thomson like solution of the scalar
configuration (\ref{Pcd7}), which by definition cannot be treated
as a small perturbation of the
metric-dilaton configuration, does not appear at all.}.\\
Yet, a similar conclusion would not be true for $\lambda_{d}
> 0$ even for $m = 0$. Indeed, in all these cases the Planck constant
$\hbar$ in Eq.(\ref{5}) becomes important and leads to the
appearance of the energy-momentum tensor for the scalar field
which is of the order of $\hbar^2$, i.e. $T^{\ M}_{\phi
N}(\hbar^2)$, hence constituting a very tiny quantity in
comparison to the energy-momentum tensor for the dilaton field,
$T^{M}_{\,N}(\hbar^0)$ (see Eqs.(\ref{row_19}) and
(\ref{row_22})), which appears on the R{\small HS} of Einstein
equations (\ref{row_3}). This means that in the first
approximation we may treat solutions of Eq.(\ref{5}) for
$\lambda_{d} > 0$ as leading to small perturbations of the
background metric (\ref{row_26}), ignoring its self-consistent
modification. In other words, the backreaction of $\phi$ is
negligible in this case.
%We will return to it later on.

\subsection{Further steps for $ k_{l} = l (l + 1) $ }
\label{further}

If the internal $\vartheta$-angle quantum number of the torus
geometry is set to $k_{l} = l (l + 1)$ then $g_{l} (\vartheta) =
P_{l} (sin\vartheta) $, $l = 0,1,2,... $, are the Lagrange
polynomians which solve Eq.(\ref{9}) and hence the second factor
on the R{\small HS} of Eq.(\ref{8}) satisfies the following
equation (see Eq.(\ref{7}))
\begin{eqnarray}
\label{10} \frac{1}{\partial \varsigma} \left( \frac{\partial
}{\partial \varsigma} h_{\tilde{\mathrm{m}}} \right) = - \left[ -
l (l + 1) + \lambda_{d} \right] h_{\tilde{\mathrm{m}}} \, .
\end{eqnarray}
The condition of continuity of functions
$h_{\tilde{\mathrm{m}}}(\varsigma)$ for $\varsigma \in [0,2 \pi )$
together with $h_{\tilde{\mathrm{m}}}(0) =
h_{\tilde{\mathrm{m}}}(2 \pi)$ give
\begin{eqnarray}
\label{11} h_{\tilde{\mathrm{m}}} (\varsigma) &=& e^{\pm i \sqrt{
- l (l +
1) + \lambda_{d} } \, \varsigma} \nonumber \\
&=& e^{i \tilde{\mathrm{m}} \varsigma} \, , \; \; \; \; \; \;
\tilde{\mathrm{m}} = 0, \pm 1, \pm 2, ... \, ,
\end{eqnarray}
where $\tilde{\mathrm{m}}$ is the internal magnetic quantum
number. Now, from the above equation we can notice that
%in the case of $k_{l} = l (l + 1)$
$\lambda_{d}$ is a quantum number which chooses the ladder of
values:
\begin{eqnarray}
\label{12} \!\!\!\!\!\!\!\!\! \lambda_{d} &=& l (l + 1) +
\tilde{\mathrm{m}}^{2} = 0,1,2,3,4,6,6,7,9,10, ...
%11,12,13,15,16,16,18,20,21,21,22,24,25,27,28,29,30,31(2),
%34,36(2),37,38,39,42(2),43,45,46(2),48,49,51(2),56(2),57,58,... \, ,
\end{eqnarray}
hence $\lambda_{d} \geq 0$. For example $\lambda_{d} =1$ in two
cases: ($l=0$ and $\tilde{\mathrm{m}}=1$) and ($l=0$ and
$\tilde{\mathrm{m}}=-1$). Let us also notice that under the
assumption of $m = 0$ the case with $\lambda_{d} = 0$ was
previously excluded~(Section~3.1)\footnote{We will see in
Section~4.3 that only the mass $m = 0$ satisfies the two physical
requirements, the first one of stability and the second one of the
proper value of the total spin of the nucleon. Hence eventually
only $\lambda_{d}
> 0$ will be possible.}.
%The choice of $m = 0$ is the very interesting physical one
%as it will be shown in Section~4.3.
%The double degeneracy of the ground state.

\subsection{The weak interaction limit}
\label{weak interaction limit}

Let us focus on the case of the weak limit which means that we are
far away ($r \gg A$) from the central core of the dylatonic field
given by Eq.(\ref{row_22}). At first let us rewrite Eq.(\ref{5})
as follows
\begin{eqnarray}
\label{Klein-Gordon2}  & & \!\!\!\!\!\!\!\!\!\!\!\!\!\!\!
\frac{A}{r^2} \left( \frac{r}{r+A} \right) \frac{\partial
u_{r}}{\partial r} - \frac{1}{r^{2}} \frac{\partial}{\partial r}
\left( r^{2}\, \frac{\partial u_{r}}{\partial r} \right) +
\left( \frac{r}{r + A} \right) \frac{L (L + 1)}{r^{2}} \, u_{r}  \\
& & \!\!\!\!\!\!\!\!\!\!\!\!\!\!\! = \frac{1}{\hbar^{2} c^{2}} \,
\biggl\{ E^{2} - m^{2} c^{4} (\frac{r}{r + A}) - \lambda_{d}
\frac{\hbar^{2} c^{2}}{d^{2}}  \left( \frac{r}{r + A} \right)^{2}
\biggr\} \, u_{r} \; . \nonumber
\end{eqnarray}
Now, asymptotically (i.e. $r \rightarrow \infty$) the first and
third term of the L{\small HS} of the above equation could be
omitted as they are of smaller order then the others. Yet, the
third term, with the orbital angular momentum $L$, will be
preserved as it is important in the angular momentum analysis.
From the point of view of the energies\footnote{The situation is
similar to the problem of a better description of hydrogen states
by the Schr{\"o}dinger equation than by the Klein-Gordon one.} of
the excited states (Section~4) it would be better in the limit $r
\rightarrow \infty$ to omit consistently all terms which are
proportional to $\frac{1}{r^2}$ since leaving only the one with
$L$ leads, in light of the omission of the first one, to the
appearance of nonphysical states.
  \\
Nevertheless let us omit the first term in
Eq.(\ref{Klein-Gordon2}) only which then, after introducing the
variable $x = A/r$, could be rewritten as follows:
\begin{eqnarray}
\label{x}  &-&   \frac{1}{A^2} x^{4} \frac{\partial^{2}
u_{r}}{\partial x^{2}} + \frac{x^2}{1 + x} \frac{L(L + 1)}{A^2}
u_{r}   \\
&=&  \frac{1}{\hbar^2 c^{2}} \left( E^{2} - m^{2} c^{4} \frac{1}{1
+ x} - \lambda_{d} \frac{\hbar^{2} c^{2}}{d^{2}} (\frac{1}{1 +
x})^{2} \right) u_{r} \, . \nonumber
\end{eqnarray}
Now, let us expand the coefficients in Eq.(\ref{x}) at point $x =
x_{0} = 0$ (i.e. $r \longrightarrow \infty$) to the second order
in~$x$. We obtain
%Finally it will appear that after neglecting the
%first derivative of the metric tensor we will obtain the
%Klein-Gordon equation which is written in the spherical coordinate
%in the asymptotically flat space.
\begin{eqnarray}
\label{13} & - &  \hbar^{2} c^{2} \left[ \frac{x^{4}}{A^2}
\frac{\partial^{2} u_{r}}{\partial x^{2}} - \left( {\it p}_{0} +
{\it p}_{1} x + {\it p}_{2} x^2 \right)
\frac{L(L + 1)}{A^2} u_{r} \right] \nonumber \\
& = &  \biggl\{ E^2 - \left( \mu_{0} + \mu_{1} x + \mu_{2} x^2
\right) m^2 c^4  - \left( \lambda_{0} + \lambda_{1} x +
\lambda_{2} x^2 \right) \lambda_{d} \left( \frac{\hbar c}{d}
\right)^{2} \biggr\} u_{r} \; ,
\end{eqnarray}
where the coefficients of the expansions and their limits for
$x_{0} = 0$ ($r \longrightarrow \infty$) are given in Appendix~B.
\\
Finally, after coming back to variable $r$ we obtain for $r \gg A$
the radial part of the Kline-Gordon-like equation, written in the
spherical coordinates as follows
\begin{eqnarray}
\label{14} - \frac{1}{r^{2}} \frac{\partial}{\partial r} \left(
r^{2} \frac{\partial u_{r}}{\partial r} \right) + \left( {\it
U}_{Rep} - {\it U}_{Attr} \right) u_{r} = \frac{ \left( E^{2} -
m_{4}^{2} c^{4} \right)}{ \hbar^{2} c^{2}} u_{r} \; ,
\end{eqnarray}
where
\begin{eqnarray}
\label{Rep-weak} \!\!\!\!\!\!\!\! {\it U}_{Rep} = \frac{L (L + 1)
+ (\frac{m c A}{\hbar})^{2} + 3 \, \lambda_{d}
(\frac{A}{d})^{2}}{r^{2}} \equiv \frac{{\cal J}^{(2)}}{r^2}  ,
\end{eqnarray}
\begin{eqnarray}
\label{Atr-weak} \!\!\!\!\!\!\!\!\!\!\!\!\! {\it U}_{Attr} =
\frac{1}{A} \frac{ \left((\frac{m c A}{\hbar})^{2} + \lambda_{d}
\, (\frac{A}{d})^{2} \right)}{r} \equiv \frac{(m_{4}^{2}
c^{4})}{(\hbar c)^{2}} \, \frac{ A}{r}
 \; ,
\end{eqnarray}
and
\begin{eqnarray}
\label{J-weak} {\cal J}^{2} = L (L + 1) + (\frac{m c
A}{\hbar})^{2} + 3 \, \lambda_{d} \, (\frac{A}{d})^{2} \; ,
\end{eqnarray}
\begin{eqnarray}
\label{m4-weak} m_{4}^{2} c^{4} = m^{2} c^{4} + \lambda_{d}
(\frac{\hbar c}{d})^{2} \; .
\end{eqnarray}
Here ${\cal J}^{2}$ can be interpreted as the square of the total
angular momentum and $m_{4}^{2}$ as the four - dimensional squared
mass of the scalar field $\phi$ in the flat Min\-kow\-skian limit
(see \cite{Dziekuje_praca z Rudnickim,Dziekuje za galaktyka}). If
the six-dimensional mass is zero ($m = 0$) then the four -
dimensional mass at spatial infinity would be solely of
kinematical origin. It is also easy to notice that neglecting the
first term of Eq.(\ref{Klein-Gordon2}) in the limit $r \rightarrow
\infty$ is at this stage indistinguishable from zeroing the first
derivative of the metric tensor\footnote{Performing physics in $r
\rightarrow \infty$ also means that we neglect in
 Eq.(\ref{Klein-Gordon2}) a "radial force"
proportional to $\partial u_{r}/\partial r$.} in $\left( \frac{-
\, r}{r+A} \right) \frac{\; \partial g^{rr}}{\partial r}
\frac{\partial u_{r}}{\partial r}$ which has its origin in the
first term at the L{\small HS} of Eq.(\ref{5}).
%(see the classical counterpart given by Eq.(\ref{row_55}) in Section~{\bf \ref{motion}}).

Now, it is interesting to notice (see Eq.(\ref{14})) that the
radial motion looks like the one dimensional motion of a particle
with a unit mass in the potential $V_{eff}$
\begin{eqnarray}
\label{efektywny} V_{eff} (r) = \frac{\hbar^{2}}{2} \left[ {\it
U}_{Rep} - {\it U}_{Attr} \right] \; .
\end{eqnarray}
The first component on the right hand side can be interpreted as a
modified centrifugal potential energy, and the second one as a
potential energy related to some central attractive force. The
existence of the second term is connected with the appearance of
the effective coupling constant which is proportional to
%which after using Eq(\ref{m4-weak}) could be expressed as follows
\begin{eqnarray}
\label{stala-weak2} \epsilon = \frac{-(\frac{m c A}{\hbar})^{2} -
\lambda_{d} \, (\frac{A}{d})^{2}}{A} = - \frac{m_{4}^{2}
c^{2}}{\hbar^{2}} A \, .
\end{eqnarray}
The value of $\epsilon$ depends both on the parameters and the
quantum numbers of the model.
%We postpone further discussion of this subject to the next Section.
  \\
After introducing $\beta = \alpha \, r$ we see \cite{Schiff-book}
that Eq.(\ref{14}) can be rewritten in the following form:
\begin{eqnarray}
\label{16} \frac{1}{\beta^{2}} \frac{\partial}{\partial \beta}
\left( \beta^{2} \frac{\partial u_{r}}{\partial \beta} \right) +
\left[ \frac{\iota}{\beta} -\frac{1}{4} - \frac{{\cal J}^{(2)} }
{\beta^{2}} \right] u_{r} = 0 \; ,
\end{eqnarray}
where
\begin{eqnarray}
\label{18} \alpha^{2} = \frac{4 \left( m_{4}^{2} c^{4} - E^{2}
\right)} {\hbar^{2} c^{2}}  \; ,
\end{eqnarray}
\begin{eqnarray}
\label{19} \iota = \frac{ - \epsilon}{\alpha} \, ,
\end{eqnarray}
with ${\cal J}^{2}$ given by Eq.(\ref{J-weak}). \\
Eq.(\ref{16}) has nearly the same formal shape as the radial
equation obtained in the Schr\"{o}dinger's model of the atom
\cite{Schiff-book}. Now, converting Eq.(\ref{18}) we obtain
\begin{eqnarray}
\label{20} E^{2} = m_{4}^{2} c^{4} - \frac{\hbar^{2} c^{2}
\alpha^{2}}{4} \; .
\end{eqnarray}
From the analysis of Eq.(\ref{16}) it can be shown that the final
solutions for $\beta = 0$ and $\beta \rightarrow \infty$ exist
only when
\begin{eqnarray}
\label{iotas} \iota = n' + j + 1 \; ,
\end{eqnarray}
where $n'$ is zero or a natural number and $j$ is a nonnegative
solution of the equation
\begin{eqnarray}
\label{sJ} j (j + 1) = {\cal J}^{(2)}
\end{eqnarray}
which is equal to
\begin{eqnarray}
\label{s} j = - \frac{1}{2} + \frac{1}{2} \left[ 1 + 4 {\cal
J}^{(2)} \right]^{\frac{1}{2}} \; .
\end{eqnarray}
Finally, using Eq.(\ref{19}) we can eliminate $\alpha$ from
Eq.(\ref{20}) and rewrite it in the following form:
\begin{eqnarray}
\label{E2-weak} \!\!\!\! E^{2} =  m_{4}^{2} c^{4} - \frac{1}{4}
\hbar^{2} c^{2} \frac{(- \epsilon)^{2}}{\iota^{2}} =  m_{4}^{2}
c^{4} \left[ 1 - \frac{1}{4} \frac{(- \epsilon)}{\iota^{2}} \, A
\right]  ,
\end{eqnarray}
where
\begin{eqnarray}
\label{iota} \iota = n'+ \frac{1}{2} + \left[ \frac{1}{4} + {\cal
J}^{(2)} \right]^{\frac{1}{2}} \,
\end{eqnarray}
and in the last equality in Eq.(\ref{E2-weak}) the expression for
$\epsilon$ given by Eq.(\ref{stala-weak2}) has been used.

%The calculations presented in this Section have been done for the
%stationary states (to the second order only) defined by a wave
%mechanics, neglecting, for fields with $m \neq 0$ their further
%interactions with the self field (played by $g_{MN}$).

At the end of this Section let us notice that the original quantum
Klein-Gordon equation given by Eq.(\ref{K-G-m}) possesses the
property of invariance according to which only the ratio
$\frac{r}{A}$ matters. We will discuss the relevant symmetry
below. Using this scale invariance, we could move from the
microscopic to the astrophysical scale. In this context the
calculations
%\footnote{In the forthcoming paper the connection between a condensate field
%(instead of $\phi$ in Eq.(\ref{rownanie}) with $m^{2} \neq 0$) and
%the Higgs field will be considered.}
point to the similarity of both the neutron and a galaxy dark
matter structure (see also \cite{Dziekuje_praca z
Rudnickim,Dziekuje za galaktyka}).

\subsection{Scale invariance}
\label{scale}

At this point the question concerning the value/s of $A$ arises.
Previously \cite{Dar,Dziekuje Ci Panie Jezu Chryste} we considered
the model given by Eq.(\ref{row_dzialanie-EH-fi}) which has the
following invariance:
\begin{eqnarray}
\label{A invariance} A \rightarrow \omega \; A \; , \;\;  r
\rightarrow \omega \; r \; , \;\;  d \rightarrow \omega \; d \; ,
\;\;  c\, t \rightarrow \omega \; c\, t \; ,
\end{eqnarray}
where $\omega$ is the parameter of the transformation. This would
mean that it is not the change of $A$ but rather of $\frac{A}{r}$
that matters. This is the invariance of the coupled Einstein
(\ref{row_3}) and Klein-Gordon (\ref{row_5}) equations but it is
not the invariance of their solution given by Eqs.(\ref{row_22})
and (\ref{row_26}). This conclusion can also be drawn from
Eq.(\ref{row_24}). Hence we notice that the massless dilatonic
field $\varphi$ is the Goldstone field. What is more, using
Eqs.(\ref{row_19})~and~(\ref{row_24}) in Eq.(\ref{row_25}) we
obtain
\begin{eqnarray}
\label{screen-in-gravity-affect} R^{r}_{r} = -\kappa_{6}
(\partial_{r} \varphi)^{2} g^{rr} = \frac{A^{2}}{2 \: r^{3} (r +
A)} \; .
\end{eqnarray}
We notice a similarity between this equation and its
electromagnetic analog\footnote{That is $\nabla^2 {\bf A} =
m_{A}^{2} {\bf A}$, where ${\bf A}$ is the electromagnetic vector
potential.}. Eg.(\ref{screen-in-gravity-affect}) is the
(anti)screening current condition in gravitation, analogous to the
screening current condition found in electromagnetism or in the
electroweak sector in the self-consistent approach
\cite{Dziekuje Ci Panie Jezu Chryste,Dziekuje_Jacek_nova_2}.\\
On the level of the equations of motion (\ref{row_3}) and
(\ref{row_5}) the theory is scale invariant unless mass $m$
enters. It would seem that invariance (\ref{A invariance}) leads
to the invariance
\begin{eqnarray}
\label{A invariance Klein} \!\!\!\!\!\!\! A \rightarrow \omega \,
A \, , \; r \rightarrow \omega \, r , \;  d \rightarrow \omega \,
d , \; c\, t \rightarrow \omega \, c \, t, \;  m \rightarrow
\frac{1}{\omega} \, m
\end{eqnarray}
of the Klein-Gordon equation (\ref{K-G-m}).
% The model (\ref{row_dzialanie-EH-fi})
% does not include any term with mass $m$.
Yet the inclusion of any term with mass $m$ into the model given
by Eq.(\ref{row_dzialanie-EH-fi}) will inevitably lead to the
gravitational coupling of the new matter with the metric tensor
part of the Lagrangian, causing the breaking of the original invariance (\ref{A invariance}). \\
Hence the conclusion emerges that there appears a relation between
the four-dimen\-sional mass $m_{4}$ and the parameter $A$. It
results from the Einstein equations which fix the value of $A$
after the self-consistent inclusion of the field $\phi$ with mass
$m$ into the  fields configuration. This means that the symmetry
connected with the scaling (\ref{A invariance}) of $A$ is broken.
Hence in Eq.(\ref{J-weak}) the dependance of ${\cal J}^{2}$ on $A$
for given $m_{4}$ is established, where $A$ is not a free
parameter. In view of the lack of a full (self-consistent)
analysis, $A$ should be estimated somehow and as we shall see this
will appear possible in the weak limit approximation.
% ($m_{4} \leftrightarrow A$)
% $(\lambda d)^{2}\leftrightarrow m_{4}^{2} \Rightarrow {\cal J}^{2}$
%because of Eqs.(\ref{m4}) or (\ref{m4-weak})
% $ A \Rightarrow {\cal J}^{2}$

\subsection{Wave mechanical limit}
\label{wave limit}

As we have proven that the global solution with $\lambda_{d}=0$
and $m=0$ does not exist (Section~\ref{global}) hence its weak
limit does not exist also. Therefore from now on we will
investigate the cases with $\lambda_{d} > 0$ only. The reason is
that, as it has been discussed in Section~\ref{global}, they only
slightly perturb the metric-dilaton configuration given by
Eqs.(\ref{row_22}) and (\ref{row_26}). At the moment let us also
suppose that $m=0$, a value which will be dynamically chosen in
the next Section.

We began with the Klein-Gordon equation (\ref{rownanie}) in which
$\phi$ is the wave function, which in quantum mechanics is usually
identified as the cause of angular momentum generators commutation
relations\footnote{Yet they may be obtained from the commutation
relations of rotation operations even before the the quantum
mechanics rules enter \cite{Sakurai}.}. In our model the square of
the total angular momentum ${\cal J}^{2} = j (j+1)$ appears
according to Eq.(\ref{J-weak}) as the combination of terms $3 \,
\lambda_{d} \, (\frac{A}{d})^{2}$ and $L (L + 1)$ which originate
from the internal two-dimensional and orbital three-dimensional
angular momenta, respectively. Yet, in the weak approximation
regime these are $j$ and $L$ which take quantum mechanical values,
therefore this means that $3 \, \lambda_{d} \,(\frac{A}{d})^{2}$
chooses quantum mechanical values as a result of the proper
addition of $j$ and $L$ only. Furthermore, since $j$, $L$ and
$\lambda_{d}$ have quantum values hence $\frac{A}{d}$ also has
this kind of values. The strange thing is that term $\frac{A}{d}$
appears as a result of the pure (field theory) classical
interaction witch originates in the background metric and now has
its part in ${\cal J}^{2}$. Next, as ${\cal J}$ manifests itself
on the arena of the tree-dimensional space as the total angular
momentum, we impose on it the usual quantum mechanical rules which
lead to different
representations for different values of ${\cal J}^{2}$.\\
This leads us to the indication of the place of wave mechanics
(quantum mechanics) which from the point of view of the
self-consistent field theory is discussed in
\cite{Dziekuje_Jacek_nova_1,Dziekuje_Jacek_nova_2}. Wave mechanics
appears as a theory for building a model of a system (e.g. an
atomic one) with the lost of self-consistency, yet it partly works
by describing its main characteristics. By "partly" I mean that to
receive a correction to the wave mechanics prediction the
self-consistency has to be reestablished. Yet this is the realm of
quantum field theory or the self-consistent field theory \cite{bib
B-K-1,Dziekuje_Jacek_nova_1,Dziekuje_Jacek_nova_2,bib J_M}. To
receive both the main characteristics and the corrections the
gravitational interactions should also be included in a
self-consistent manner where the linear part of field fluctuations
would be described by the quantum field theory. For the detailed
discussion of the place of the quantum field theory see \cite{bib
B-K-1}.

\section{Numerical example: A neutron and its excited states}
\label{numerical example}

Although the (internal) spin wave function could not have the
$(\Theta, \Phi)$ space representation for the physical half-spin
state, yet it does not eliminate motion as the origin of the spin
in some kind of the classical space at all. Besides unknown,
mainly three possibilities could be taken into account. The first
one states that the parameters of the spin representations are
"seen" in the interactions only without further inner explanation.
The second one exists by extending the space by the Grassmann
coordinates \cite{Mannheim}. The third one, considered in this
paper, constructs the spin of the particle from the motion in the
geometry of the inner space. The half-integer spin is then, as any
other, just the numerical result.\\
The other spin representations of fields could have the similar
origin (see Section~1.1). Yet there is a reason to consider them
as having the different dimensions of their inner spaces. The
prove goes via the Fisherian statistical analysis performed by
Frieden \cite{Frieden} and the conclusion is that the dimension of
the inner space depends on the species of the field of the rank
$\aleph$ which is the dimension of the sample collected by the
field alone during its sampling of the space-time. For the complex
scalar field the dimension of the extra parametric space is equal
to $\aleph=2$. For the other fields this dimension is bigger, e.g
$\aleph=4$ for the electromagnetic field and $\aleph=8$ for the
Dirac one. Hence, as it has been mentioned previously, the
four-dimensional space-time could be extended simultaneously to
the variety of the multi-dimensional ones.

\subsection{Neutron}
\label{neutron}

Let us make some numerical adjustments for the parameters of the
presented model ($\aleph=2$). For this purpose we begin with the
state which possesses the following value of the internal quantum
number (see Sections~3.1-2):
\begin{eqnarray}
\label{parameters} \lambda_{d} = 1 \; ,
\end{eqnarray}
choosing the following value of the four dimensional mass:
\begin{eqnarray}
\label{m4_first} m_{4} \approx 952.893194 \;\, MeV/c^{2} \; ,
\end{eqnarray}
which is a little bit bigger than the observed mass of the neutron
$939.56536 \; MeV/c^{2}$. Solving Eq(\ref{m4-weak}) we obtain
\begin{eqnarray}
\label{dl} d  = d_{l} \approx 2.071\;\, 10^{-16} \; {\rm m} \;  =
0.2071 \; {\rm fm}
\end{eqnarray}
as the limit value, i.e. from Eq(\ref{m4-weak}) we see that for $d
> d_{l}$ we obtain $m^{2} > 0$ and for $d < d_{l}$ we obtain
$m^{2} < 0$ which is a tachion. For the limit value $m^{2} = 0$
the second term of ${\cal J}^{(2)}$ at the R{\small HS} of
Eq.(\ref{J-weak}) is equal to
\begin{eqnarray}
(\frac{m c A}{\hbar})^{2} = 0 \; .
\end{eqnarray}

Now, it is interesting to notice that the R{\small HS} of
Eq.(\ref{J-weak}) might be interpreted as $j (j + 1)$ (see
Eq.(\ref{sJ})), which means that
\begin{eqnarray}
\label{j-row-weak-2} {\cal J}^{(2)} &=& L (L + 1) + (\frac{m c
A}{\hbar})^{2} + 3 \, \lambda_{d} \, (\frac{A}{d})^{2}  \nonumber \\
&=& j(j + 1) \; .
\end{eqnarray}
After taking $L=0$ and $j = 1/2$ for the neutron the equation on
the relative value of $A$ and $d$ is obtained from the above
equation. Moreover, we see that Eq.(\ref{j-row-weak-2}) gives us
\begin{eqnarray}
\label{A-value-weak} \frac{A}{d} = 1/2 \;\;\;\;\;\; {\rm for} \;\;
d = d_{l} \; , \;\;\; ( m^{2} = 0 ) \, .
\end{eqnarray}
Now, according to Eqs.(\ref{E2-weak}) and (\ref{stala-weak2}) we
obtain for $m=0$ and $\frac{A}{d}$ as in Eq.(\ref{A-value-weak})
the following equation
\begin{eqnarray}
\label{E-weak} \!\!\!\!\!\!\! E^2 =  m_{4}^2 c^{4} \left[ 1 -
\frac{1}{4} \frac{(- \epsilon A)}{\iota^{2}} \right] = m_{4}^2
c^{4} \left[ 1 - \frac{1}{36} (\frac{3/2}{\iota})^2 \right]
\end{eqnarray}
and using in this equation $m_{4}$ from Eq.(\ref{m4_first}) we
find that the energy of the neutron in the ground state ($n' = 0$,
$j = 1/2$ and $\iota = 3/2$, see Eq.(\ref{iotas})) is equal to
\begin{eqnarray}
\label{E obserwowana} E = 939.565 \; MeV/c^{2}
\end{eqnarray}
which is the observed mass of the neutron. From the second term of
Eq.(\ref{E-weak}) it results that the gravitational correction to
the energy of the nucleon in the ground state ($n' = 0$ and $j =
1/2$ and $\iota = 3/2$), i.e. the gravitational bounding energy,
is equal to
\begin{eqnarray}
\label{E-value-weak} \Delta E = E - m_{4} c^2  \approx - 0.0139867
\; m_{4} c^2 \; .
\end{eqnarray}
Finally, let us recall that the solution for $\lambda_{d} =1$ has
according to Eqs.(\ref{11}) and (\ref{12}) the double
degeneration: ($l=0$ and $\tilde{\mathrm{m}}=1$ - which we
interpret as the spin up solution) or ($l=0$ and
$\tilde{\mathrm{m}}=-1$ - which we interpret as the spin down
solution). This assignment is very natural but by its own it does
not lead to the conclusion that the particle is the neutron. The
calculation of the energy of the ground state of a field having
the "quantum" numbers characteristic for the neutron entitles only
to the conclusion that the formalism could describe the neutron.
Hence we see that the model points to the origin of the spin of
the nucleon as connected with the geometry of the internal
manifold and with motion inside its two compactified parametric
dimensions. Moreover, it indicates that gravitational effects give
a visible change of the spectrum of the levels in the proposed
model of the nucleon. \\
As it has been shown, the background gravity-dilatonic
configuration given by Eqs.(\ref{row_22}) and (\ref{row_26}) is
stable (see Section~2.1). Also the ground state solution obtained
in Section~\ref{implication} for the scalar field $\phi$ is the
stationary one. Moreover, it perturbs the background configuration
slightly only  (see Section~3.1) and in the discussed weak
approximation we could argue that the whole configuration of the
fields, $g_{MN}$, $\varphi$ and $\phi$, is stable. Finally, for a
particle with another value of its spin the dimension $\aleph$ of
the inner parametric space is different, yet the conclusion, if
the relevant calculations are performed, would be similar.

\subsection{The value of $m = 0$}
\label{value m}

Let us still consider one particular state (e.g. neutron) with
$m_{4}^{2}$ {\it fixed}, i.e. $m$ and $d$ may vary leaving
$m_{4}^{2}$ unchanged.  Now, with the use of Eq.(\ref{m4-weak}) we
notice that Eq.(\ref{j-row-weak-2}) can be rewritten as follows
\begin{eqnarray}
\label{J2-m42} {\cal J}^{(2)} = j(j + 1)  = L (L + 1) + m_{4}^{2}
c^{4} (\frac{A}{c \, \hbar})^{2} + 2 \lambda_{d} (\frac{A}{d})^{2}
\end{eqnarray}
hence we see that ${\cal J}^{(2)}$ for $\lambda_{d} > 0$ is not
proportional to $m_{4}^{2} c^{4}$ (no matter what the integer $L$
is) and it is worth noting that to keep ${\cal J}^{(2)}$ equal to
$j (j+1)$ for $j$ equal to the multiplicity of $1/2$, it is
necessary to have $m=0$ (hence $m_{4}^{2} c^{4} = \lambda_{d}
(\frac{\hbar c}{d})^{2}$) without any variation from this
particular value of $m$.

\subsection{Excited states}
\label{excited}

As in Section~\ref{neutron} the value of $m_{4}$ for the neutron
has been chosen to fit the value of the observed energy $E$ of the
neutron hence it is important to calculate the predictions for
values of the energy of its excited states. Let us make the
following assumption on the relation between the total angular
momentum number and the orbital one:
\begin{eqnarray}
\label{j-L} j = L + \frac{1}{2} .
\end{eqnarray}
For the mass $m = 0$ we may notice from Eq.(\ref{j-row-weak-2})
that the characteristic radius $A$ has to change with $j$, $L$ and
$\lambda_{d}$ in the following way
\begin{eqnarray}
\label{A} A = \frac{\, d \; {\sqrt{j\,\left( 1 + j \right)  -
L\,\left( 1 + L \right) }}}{{\sqrt{3}}\,{\sqrt{\lambda_{d}}}}
\;\;\;\; \;\;\; {\rm for} \;\;\;\; m = 0 \; ,
\end{eqnarray}
where $d$ is kept equal to $d_{l} = 0.2071$ $\left[ {\rm fm}
\right]$ as in Eq.(\ref{dl}). Then, with $m_4$ given for each
particular state by Eq.(\ref{m4-weak}), the energy $E$ is
calculated according to Eq.(\ref{E2-weak}). The numerical values
of the energy $E$ for the ground state ($j=\frac{1}{2}$,
$\lambda_{d} = 1$) and excited states of the neutron are given in
the Table~\ref{table} below.\\
Except for the ground state (which has been chosen to fit the
neutron observed energy) all other energies are not ideally equal
to the observed ones \cite{particle_review}. Nevertheless this
might not be a big shortcoming of the model since we use the weak
approximation only. This statement appears to be sensible,
especially in the light of the agreement of the order of the
discrepancies between energies of the states belonging to the same
column and different rows, which are of the order found in the
experiment. As the spacial configurations for the first column in
the Table~1 are the most simple ones ($L=0$) hence the feeling is
that only they should lie in the neighborhoods of the true states
(see also the discussion after Eq.(\ref{Klein-Gordon2})).
\begin{table}[here]
\caption{Values of energies $E$ MeV for the ground state and
excited states of the neutron as the function of the total angular
momentum number $j$ and internal quantum number $\lambda_{d}$. For
the neutron itself $j=\frac{1}{2}$ and $\lambda_{d} = 1$. The last
two columns present the gravitational bounding energy $\Delta E$
MeV and the ratio $A/d$ for the neutron and its excited states
with $j=1/2$ and $L=0$. All states in the Table~1 have $m=0$ and
$d = d_{l} = 0.2071$ fm (see the text).}
\label{table}       % Give a unique label
\begin{center}
\begin{tabular}{|c|c|c|c|c|c|c|c|c}
  \hline
    $\!\! \lambda_{d}\!\!$ &  $\!\!j=$$1/2\!\!$ &  $\!\!3/2\!\!$ &  $\!\!5/2\!\!$ & $\!\!7/2\!\!$ &
 $\!\!9/2\!\!$ & $\!\!\Delta E (j=1/2)\!\!$ & $\!\!A/d$ $(j=1/2)\!\!$\\
  \hline
  1 & {\bf 939.565} & 941.71 & 943.938 & 945.512 & 946.638 & {\bf -13.3278} & {\bf 0.5} \\
  2 & {\bf 1328.75} & 1331.78 & 1334.93 & 1337.16 & 1338.75 & {\bf -18.8484} & {\bf 0.3536} \\
  3 & {\bf 1627.37} & 1631.09 & 1634.95 & 1637.67 & 1639.63  & {\bf -23.0845} & {\bf 0.2887} \\
  4 & {\bf 1879.13} & 1883.42 & 1887.88 & 1891.02 & 1893.28  & {\bf -26.6557} & {\bf 0.25} \\
  6 & {\bf 2301.46} & 2306.71 & 2312.17 & 2316.02 & 2318.78  & {\bf -32.6464} & {\bf 0.2041} \\
  %7 & 2485.86 & 2491.53 & 2497.43 & 2501.59 & 2504.57  & -35.2621 & 0.188982 \\
  %9 & 2818.7 & 2825.13 & 2831.81 & 2836.54 & 2839.91   & -39.9835 & 0.166667 \\
  %10 & 2971.17 & 2977.95 & 2984.99 & 2989.97 & 2993.53 & -42.1463 & 0.158114 \\
    \hline
    & $L=$ 0       & 1       & 2       & 3       & 4  & $L=0$ & $L=0$\\
  \hline
\end{tabular}
\end{center}
\end{table}

\subsubsection{The gravitational bounding energy}
\label{bounding_energy}

Additionally, in the two last columns of the Table~1, the values
of the gravitational bounding energy
\begin{eqnarray}
\label{bounding} \Delta E =  E - m_{4} c^{2} = m_{4} c^{2} \left[
\sqrt{1 - \frac{1}{4} \frac{(- \epsilon A)}{\iota^{2}}} -1 \right]
\; ,
\end{eqnarray}
for the states with $j=\frac{1}{2}$ and $L=0$ are given. Here,
according to Eqs.(\ref{stala-weak2}) and (\ref{m4-weak}) the
effective coupling constant $\epsilon$ and the the four -
dimensional squared mass $m_{4}^{2}$ are for $m=0$ equal to
\begin{eqnarray}
\label{stala i m4-weak dla m 0} \epsilon = - \frac{m_{4}^{2}
c^{2}}{\hbar^{2}} A \, \;\;\;\; {\rm and} \;\;\;\;\; m_{4}^{2}
c^{4} = \lambda_{d} (\frac{\hbar c}{d})^{2} \;\;\;\; {\rm for}
\;\;\; m = 0 \; ,
\end{eqnarray}
where $A$ is given by Eq.(\ref{A}).\\
The important result is that for fixed $j$ and $L$, the increase
of the energy $E$ of the configuration causes the system to become
more tightened (as the ratio $A/d$ decreases) and simultaneously
its gravitational bounding energy bigger. As an example, for the
neutron and its excited states (with $j=1/2$ and $L=0$) this
result has been presented in the first and in the two last columns
of the Table~1 (in bold). \\
The model could be applied to the other known particles also, yet
the present calculations show that the best agreement is achieved
for the neutron and its excited states. For example, for the
neutral meson $\rho_{o}$ taken as the ground state and the first
five of its excited states, for all of which we choose $j = 1$,
$n' = 0$, $\iota = 2$ and $L=0$, we obtain the energies equal to
$1454.5$, $2056.97$, $2519.26$, $2909.0$, $3562.78$, $3848.24$ MeV
for $\lambda_{d}$ equal to 1,2,3,4,6,7, respectively. So, in this
case we notice the worse matching with the experimental values
\cite{particle_review} than in the case of the neutron. The
explanation of this fact might be that the gravitational factor in
the structure of the neutron, which is the lowest known barionic
neutral ground state, is most decisive than in the case of other
neutral barions.

\subsubsection{The problem with the degeneracy}

The principal problem seems to be connected with the ladder of the
energy state values in each particular row since in each row few
of them or may be only one are physical. This problem might be
connected with the zeroing of the term with the first derivative
of the metric tensor and hence with the removal of the first
derivative of the radial wave function $u_{r}$ on passing from
Eq.(\ref{Klein-Gordon2}) to Eq.(\ref{x}).
% which is equivalent to the addition of the correct inertial force.
Unfortunately, after this zeroing, the weak limit approximation
was invented, leading effectively to the introduction of an extra
force term (which is equal to the neglected one with the opposite
sign) connected with the mathematics of the weak approximation
only. The addition of this unphysical extra force into the system
results in appearance of row ladders of the unphysical states. Yet
the unphysical differences between states of the ladder in one
particular row with the established value of $\lambda_{d}$ are two
order of magnitude smaller than the physical ones from the
columns. This agrees with the fact that the differences in one row
appear as the result of the change in the value of $L$. Now,
according to Section~\ref{weak interaction limit} the orbital
angular momentum $L$ survived with the third term $\left(
\frac{r}{r + A} \right) \frac{L (L + 1)}{r^{2}} \, u_{r}$ of
Eq.(\ref{Klein-Gordon2}). Yet it survived inconsistently because
the first term $\frac{A}{r^2} \left( \frac{r}{r+A} \right)
\frac{\partial u_{r}}{\partial r}$ of this equation, also
proportional to $\frac{1}{r^{2}}$, has been neglected. Since the
third term of Eq.(\ref{Klein-Gordon2}) is, for $r \rightarrow
\infty$, of the order smaller than the ones which remained
together with it in Eq.(\ref{x}) (and consequently in
Eq.(\ref{14})) therefore the energy differences in a particular
row are also smaller than in a particular column. The analysis of
Eqs.(\ref{14})-(\ref{Atr-weak}) is in agrement with the remarks
above, i.e. the change of $L$ in ${\it U}_{Rep} \sim
\frac{1}{r^{2}}$ is connected with the small energy differences in
a particular row whereas the change of $\lambda_{d}$ in ${\it
U}_{Attr} \sim \frac{1}{r}$ is connected with the big energy
differences in a particular column of the Table~\ref{table}. As we
noticed in Section~\ref{weak interaction limit} the term with $L$
has been left for the sake of the angular momentum considerations.

\section{Conclusions}
\label{concl}

Till now there does not exist established experimental evidence
for multi-dimensionality of the world and/or our understanding of
potential manifestations of higher dimensions is too poor. Yet at
the same time there exist in the literature multiple references
indicating that the pursued effects of extra dimensions could have
already been noticed earlier, both in the astrophysical
\cite{Wesson-Overduin,Biesiada_Malec}, \cite{Dar,Dziekuje Ci Panie
Jezu Chryste,Dziekuje_praca z Rudnickim,Dziekuje za galaktyka} and
in elementary particle scale \cite{Gr-Sch-Kaku},
\cite{Dziekuje_Jacek_nova_1,extra_dimensions}, contrary to the
standard expectations that extremely high energies are necessary
to probe higher dimensions. The renewed interest in the
Kaluza-Klein theories, to which the presented six-dimensional
model belongs, stems from the fact that multidimensional analogues
of general relativity are able, among others, to generate the four
dimensional mass out of the interaction which proceeds both from
the four-dimensional world and from the internal space dimensions
(here two) with currents of the matter which modify masses in the
four-dimensional world (see also \cite{Dziekuje za galaktyka}).
Our approach to the modelling of a system, e.g. of one particle
(here neutron), stems from the hope that its structure might be
described by the dynamics in the six-dimensional space-time, and
especially that its spin and, at least partly its inertia have
their origin in the internal two-dimensional space. This is the
reason for perceiving six-dimensional (and more generally,
multi-dimensional\footnote{For example, in the case of the
electromagnetism for which $\aleph = 4$, the space-time would be
eight-dimensional.}) theories as still attractive for
understanding both fundamental interactions and the more
sophisticated considerations which are related to the problem of
Mach's principle.

To fulfil this aim the present paper examines the implementation
of the geometrical properties of the previously \cite{Dar} worked
out six-dimensional Kaluza-Klein type model which manifest
themselves in possible observational consequences in the realm of
one particle physics, e.g. having an impact on the structure of
neutron and its excited states. The obtained ground state
solution, here neutron, appeared to have two spinorial degrees of
freedom as the total angular momentum with $j=1/2$ goes (for the
main internal angular quantum number $\lambda_{d} = 1$) with two
internal magnetic quantum number possibilities,
$\tilde{\mathrm{m}}=1$ or $\tilde{\mathrm{m}}=-1$. In Section~1.1,
the statistical Fisherian reasons for the six-dimensionality of
the space-time are given. The fundamentals of the model are
presented in Section~\ref{geometry} where the static spherically
symmetric "reference" back-ground solution of the six-dimensional
Einstein equations coupled with the Klein-Gordon equation with the
dilatonic field as the basic one and the gravitational field as
the self-field, is recalled \cite{Dar}. A more detailed discussion
of its properties is presented in \cite{Dar,Dziekuje Ci Panie Jezu
Chryste,Dziekuje_praca z Rudnickim,Dziekuje za galaktyka}. The
metric tensor part of the "reference" solution is, in a sense,
analogous to the familiar four-dimensional Schwarzschild solution
but yet fundamentally different since it is horizon free. In
Section~\ref{stability} the stability of the "reference"
self-consistent solution is shown. Section~\ref{implication} is
devoted to the derivation of the weak limit approximation of the
general wave mechanical (quantum mechanical) approach which is
defined in the context of losing (here gravitational)
self-consistency. So, a new scalar field has been added to the
original "reference" configuration of fields and the non
self-consistent case for its Klein-Gordon equation is defined.
Here the main purpose is to find the solution of this Klein-Gordon
equation and the reason for doing this is twofold. On one hand it
is connected with the derivation of the energy of states. On the
other hand with the indication of the spin origin of the
configuration as the manifestation of both the geometry of the
internal two-dimensional space and the kinematics of the fields
inside it. In Section~\ref{global} the problem of the departure
from the (gravitational) self-consistent calculations of the
metric and other fields of the configuration has been discussed.
The lost of the self-consistency is connected with neglecting the
perturbation term (of the total energy-momentum tensor) which is
proportional to $\hbar^2$. In Section~\ref{numerical example} the
implementation of the model to the description of neutron and its
excited states has been performed, including the derivation of
their energy. Finally, the discussion of the consequences of
zeroing of the first derivative term in the radial Klein-Gordon
equation has been given in Section~\ref{excited}.

The question arises on the status of the presented model of the
neutron. As discussed previously in Sections~\ref{implication} and
\ref{numerical example} it is interpreted as the wave mechanical
departure from the fully (i.e. including gravitation)
self-consistent field theory understood in accordance with the
Schr\"{o}dinger's substantial interpretation of the wave function
\cite{Dziekuje_Jacek_nova_1,Dziekuje_Jacek_nova_2}. The wave
mechanical approach means that the self-consistency is lost. Yet
there is hope that a fully self-consistent model which
incorporates all the necessary fields and interactions will result
not only in the refinement of the proposed model of the neutron
and its excited states but also in the construction of better
models of all other elementary matter particles. Yet this is the
difficult task to perform \cite{bib
B-K-1,Dziekuje_Jacek_nova_1,Dziekuje_Jacek_nova_2,bib J_M}. \\
In today's physics mainly the wave mechanical (weak approximation
case) and quantum field theory (linear) regimes are perceived. Yet
self-consistent exceptions are also quoted, developed only for the
description of the (mainly linear) fluctuations of the matter wave
function \cite{bib B-K-1}. The concept which lies behind this
approach is not the quantum theory of interactions with quantum
gravity included \cite{Ashtekar} with the second quantization
procedure for a set of aggregated quanta. But, it is rather of a
relativistic self-consistent field theory origin which includes
all interactions, with gravity in this number. The consistency of
this approach is guaranteed on the theoretical basis which is
stronger than the mean-field theory concept as both wave mechanics
and classical field theories can be understood as having the same,
Fisher-information origin with finite $\aleph$
\cite{Frieden,Dziekuje_informacja}.\\
Finally, according to the calculation of this paper, Mach's
principle has to be modified. Since the mass of a particle comes
both from external and internal dimensions (where the internal
part of this mass is purely of the kinetic origin) hence it is
essential that the final mass of the particle should be calculated
in the self-consistent formalism. It might be done in accord with
a broadened version of Thrirring and Einstein analysis of Einstein
geometrodynamics or in accord with an effective gravity theory of
the Logunov type \cite{Denisov-Logunov}. In both cases the
internal contribution to the mass should appear as the result of
the self-consistent calculations. Interestingly the same
conclusions were drowned by Frieden \cite{Frieden} from the
statistical method of estimating the physical models and just
recently proved in \cite{Dziekuje_informacja}, where the
entanglement between the kinematical and structural degrees of
freedom has been rigorously established. Now, because the mass has
its origin in the dynamics and the kinetic motion of the
(extended) field of every elementary particle inside the internal
space, we obtain an elementary property according to which every
particle is a continuously extended object. \\
Yet, the presented model is obviously not the final one. The
reason is twofold. At first, from its beginning it was formulated
for the neutral particle only. The inclusion of e.g. the electric
or effective weak charges, which would be necessary, changes the
results\footnote{For the self-consistent analysis in the classical
counterpart of the electroweak model see
\cite{Dziekuje_Jacek_nova_2}.}. At second, the first particle in
every chosen ladder of states should really be the ground state of
the configuration of all fields which are taken into account to
describe self-consistently both the ground state and excited
states. Hence, further calculations should incorporate more
realistic shapes of both the main charge densities and their
fluctuations for the extended basic matter sources to which proper
self-fields are coupled. These shapes should follow both from the
Einstein equations coupled to e.g. Klein-Gordon-Maxwell (effective
Yang-Mills) or Dirac-Maxwell (effective Yang-Mills) set of
equations for charge densities and fluctuations simultaneously, as
it is required for the self-consistent models
\cite{Dziekuje_Jacek_nova_2}. It means that from the mathematical
point of view, a matter particle seems to be a self-consistent
solution of field equations for the basic fields (including
fluctuations) and their self-fields which are involved in the
description of this particle. The presented model indicates the
place where the weak, wave mechanical limit of the theory lies and
is a step towards the construction of the self-consistent
formalism of the classical theory of one, continuously extended,
elementary particle.

\section*{Acknowledgments}
This work has been supported by L.J.Ch..\\
Special thanks to Barbara Mierzy{\'n}ska for the correction of the
text. This work has been also supported by the Department of Field
Theory and Particle Physics, Institute of Physics, University of
Silesia, the project "{\it Teoretyczne i fenomenologiczne badanie
Modelu Standardowego i jego uog{\' o}lnie{\'n}~}", No:
BS-03-0507-022-07 and by the Junta de Andaluc{\' i}a project FQM
437.

\section*{Appendix A: The components of the Ricci tensor of the background
metric}

Using Eqs.(\ref{row_7})-(\ref{row_8}), we can calculate the
components of the Ricci tensor. The nonvanishing components are
\cite{Dar,Dziekuje Ci Panie Jezu Chryste}
\begin{eqnarray}  \label{row_10}
R^{t}_{t} = \left( 4 \: \varrho^{2} r \nu^{\prime}+ 4 \: r^2
\varrho^{\prime}\varrho \nu^{\prime} - r^{2} \varrho^{2}
\mu^{\prime}\nu^{\prime} + r^{2} \varrho^{2} (\nu^{\prime})^{2} +
2 \: r^{2} \varrho^{2} \nu^{\prime\prime} \right) (4 \: e^{\mu}
r^{2} \varrho^{2})^{-1}
\end{eqnarray}
\begin{eqnarray}  \label{row_11}
R^{r}_{r} &=& \left(- 4 \: \varrho^{2} r \mu^{\prime} - 4 \: r^2
\varrho^{\prime}\varrho \mu^{\prime} - r^{2} \varrho^{2}
\mu^{\prime}\nu^{\prime} +  r^{2} \varrho^{2} (\nu^{\prime})^{2} +
8 \: r^{2} \varrho \varrho^{\prime\prime} + \right. \nonumber \\
&+& \left. 2 \: r^{2} \varrho^{2} \nu^{\prime\prime} \right)
\left( 4 \: e^{\mu} r^{2} \varrho^{2} \right)^{-1}
\end{eqnarray}
\begin{eqnarray}  \label{row_12} \!\!\!\!\! R^{\Theta}_{\Theta} =
R^{\Phi}_{\Phi} = \left( - 4 \: e^{\mu} \varrho^{2} + 4 \:
\varrho^{2} + 8 r \varrho \varrho^{\prime} - 2 \: r \varrho^{2}
\mu^{\prime} + 2 \: r \varrho^{2} \nu^{\prime} \right) \left( 4 \:
e^{\mu} r^{2} \varrho^{2} \right)^{-1}
\end{eqnarray}
\begin{eqnarray}  \label{row_14} \!\!\!\! R^{\vartheta}_{\vartheta} =
R^{\varsigma}_{\varsigma} = \left( 8 \: \varrho r \varrho^{\prime}
+ 4 \: r^2 (\varrho^{\prime})^{2} - 2 \: r^{2} \varrho
\varrho^{\prime}\mu^{\prime} +  2 \: r^{2} \varrho
\varrho^{\prime}\nu^{\prime} + 4 \: r^{2} \varrho
\varrho^{\prime\prime} \right) \left( 4 \: e^{\mu} r^{2}
\varrho^{2} \right)^{-1}
\end{eqnarray}

\section*{Appendix B: The weak interaction limit coefficients}

The coefficients of expansions of the radial equation (\ref{13})
and their limits for $x_{0} = 0$ ($r \longrightarrow \infty$) are
equal to
\begin{eqnarray}
\label{p} \!\!\!\! {\it p}_{0} = - \frac{x_{0}^{3}}{(1 +
x_{0})^{3}} = 0 \, , \;\;\;\;\; {\it p}_{1} =  \frac{3 x_{0}^{2} +
x_{0}^{3}}{(1 + x_{0})^{3}}  = 0 \, , \;\;\;\;\;\;\; {\it p}_{2} =
\frac{1}{(1 + x_{0})^{3}}  = 1 \, ,
\end{eqnarray}
\begin{eqnarray}
\label{m}  \mu_{0} =  \frac{1 + 3 x_{0} + 3 x_{0}^{2}}{(1 +
x_{0})^{3}}  = 1 , \;\; \mu_{1} =  - \frac{1 + 3 x_{0}}{(1 +
x_{0})^{3}}  = -1 \, , \;\;  \mu_{2} =  \frac{1}{(1 + x_{0})^{3}}
= 1 \, ,
\end{eqnarray}
\begin{eqnarray}
\label{ld}  \lambda_{0} =  \frac{1 + 3 x_{0} + 5 x_{0}^{2}}{(1 +
x_{0})^{4}} = 1 , \;\; \lambda_{1} =  - \frac{1 + 7 x_{0}}{(1 +
x_{0})^{4}}  = -1  \, , \;\; \lambda_{2} =  \frac{3}{(1 +
x_{0})^{4}} = 3 \, .
\end{eqnarray}

\end{document}